     \def\q{\hbox{\raisebox{.2ex}{$q$}}}
     
     \def\sdag{\scriptscriptstyle \dag}
     \def\sL{\scriptscriptstyle L}
     \def\sR{\scriptscriptstyle R}
     \def\ps{\hbox{\raisebox{.2ex}{$p$}}\mbox{\hspace{-0.9ex}}/}

\documentstyle{article}

\font\tenrm=cmr10

\font\elevenbf=cmbx10 scaled\magstep 1
\font\elevenrm=cmr10 scaled\magstep 1
\font\elevenit=cmti10 scaled\magstep 1

\renewenvironment{thebibliography}[1]
 { \elevenrm
   \begin{list}{\arabic{enumi}.}
    {\usecounter{enumi} \setlength{\parsep}{0pt}
     \setlength{\itemsep}{3pt} \settowidth{\labelwidth}{#1.}
     \sloppy
    }}{\end{list}}

\begin{document}
\begin{flushright}
\begin{tabular}{l}
HUPD-9220\hspace{0.5cm}\\
December 1992
\end{tabular}
\end{flushright}
\vglue 0.9cm
\begin{center}
{{\Huge Dynamical CP Violation\\
         in Composite Higgs Models\\}
\vglue 1.9cm
{\Large S.~Hashimoto,\hspace{-0.1cm}%
\renewcommand{\thefootnote}{\fnsymbol{footnote}}
\footnote{Supported in part by Monbusho Grant-in-Aid under the
 contract No.040011}
\hspace{0.05cm} T.~Inagaki and T.~Muta
\footnote{Supported in part by Monbusho Grant-in-Aid
for Scientific Research (C) under the contract No. 04640301}\\}
\baselineskip=13pt
\vglue 0.55cm
{\elevenit Department of Physics, Hiroshima University\\}
\baselineskip=12pt
\vglue 0.15cm
{\elevenit Higashi-Hiroshima, Hiroshima 724, Japan\\}
\vglue 5.2cm

{\Large ABSTRACT}}
\end{center}
\vglue 0.2cm
{\rightskip=3pc
 \leftskip=3pc
 \elevenrm\baselineskip=16pt
 \noindent
 The dynamical origin of the CP violation in electroweak theory
 is investigated in composite Higgs models.
 The mechanism of the spontaneous CP violation proposed in other
 context  by Dashen is adopted to construct simple
 models of the dynamical CP violation.
 Within the models the size of the neutron electric dipole moment is
 estimated and
 the constraint on the $\varepsilon$-parameter in K-meson decays is
 discussed.
\vglue 2.2cm}

\newpage

\renewcommand{\thesubsection}{\Roman{subsection}.}
\subsection{INTRODUCTION}
\stepcounter{section}
\vglue 0.1cm
\baselineskip=18pt
\elevenrm
\hspace*{\parindent}
The CP violation is described by phases appearing in the
 Kobayashi-Maskawa matrix\cite{KM}
 in the standard theory of quarks and leptons.
 The CP violating phases are introduced only when the number of the
 quark-lepton
 generations is equal to or greater than three.
 In other words the reason why we have the CP violation in nature is
 that we have three
 generations of quarks and leptons.
 The CP violating phases are partially determined by experimental
 data in the neutral K-meson system.
 The prediction for the neutron electric dipole moment\cite{NEDM}
 based on the Kobayashi-Maskawa CP violating
 phases (KM phases) is extremely small and is well below the
 experimental lower bound\cite{NEDE}.
 Thus the standard theory with Kobayashi-Maskawa CP violation is
 consistent with the present experimental situation.\\
\hspace*{\parindent}The KM phases are introduced as free parameters
 in the standard theory.
 From the point of view of the fundamental theory of quarks and
 leptons this situation is not satisfactory and we would like to see
 where is the theoretical origin of the KM phases describing the CP
 violation.\\
\hspace*{\parindent}One of the possibilities to explain the KM phases
 by the more fundamental origin is to introduce the complex vacuum
 expectation values for the Higgs field as discussed by Weinberg more
 than a decade ago\cite{SCP}.
 In this approach it is required to have at least three Higgs doublets
 in order to interprete the full KM phases.
 This mechanism suggests that the spontaneous electroweak symmetry
 breaking has something to do with the origin of the CP violation.\\
\hspace*{\parindent}Pushing forward this idea we are naturally led to
 the composite Higgs
 models where the Higgs field is replaced by a composite system of
 fundamental fermions.
 There are a variety of the composite Higgs models including the
 technicolor model\cite{TECH}, top-condensation model\cite{TOPC,TCOL},
 fourth-generation model\cite{FG} and color-sextet quark
 model\cite{SQ}.
 In the composite Higgs models the CP violation may occur if the
 complex vacuum expectation value would result for the composite
 field $\bar{\psi}\psi$ with fundamental fermion $\psi$.
 The realization of such circumstance has once been suggested long
 time ago by Dashen in other context\cite{ODCP}.\\
\hspace*{\parindent}The idea of Dashen will be recapitulated in
 the next section and will be applied straightforwardly to the
 composite Higgs models.
 Eichten, Lane and Preskill\cite{DCP1} have adopted Dashen's idea in
 the technicolor model to elucidate the mechanism of the dynamical
 CP violation.
 In their paper the general framework of generating the dynamical
 CP violation was presented and some physical consequences were
 pointed out.
 Later Goldstein\cite{DCP2} has reconsidered the problem and
 constructed a model of the dynamical CP violation with two quark
 and techniquark doublets.
 This model, however, fails to give rise to the CP-violating phase
 unless one introduces extra leptons or one assumes an existence of
 the strong CP violation in the technicolor sector.\\
\hspace*{\parindent}In the present paper we would like to construct
 some simple examples of the dynamical CP violation in the composite
 Higgs models.
 In our models we assume the presence of two flavors of up(down)-type
 extra fundamental quarks and three flavors of up(down)-type ordinary
 quarks.
 We start with the Lagrangian with flavor symmetry
 (i.e. all fermions massless)
 in which a nonvanishing vacuum expectation value develops for the
 composite field $\bar{\psi}\psi$ with $\psi$ the fundamental fermion.
 To this Lagrangian we add flavor-symmetry breaking terms to realize
 the quark mass hierarchy.
 We consider transformations which mixes the flavors of quarks.
 We find a special solution for the transformations which gives the
 true vacuum with the proper direction.
 According to this special solution the CP violating terms are
 generated in the flavor-symmetry breaking part of the Lagrangian.\\
\hspace*{\parindent}The main purpose of our argument is to show the
 usefulness of the Dashen mechanism for the dynamical CP violation
 in a transparent way.
 Our model is too simple to explain the KM phases practically and
 should be elaborated to reproduce the standard theory as a
 low-energy effective theory.
 If our model has something to do with the nature, it has to be
 consistent with the existing experimental observations.
 Thus we calculate the contribution in our model to the electric
 dipole moment of the neutron and the $\varepsilon$-parameter in
 K decays.
 Both quantities are found to be consistent with the experimental
 data if the cut-off $\Lambda$ existing in the model is larger than
 $800$ TeV which is consistent with the cut-off set by the FCNC
 restriction\cite{FCNC}.\\
\hspace*{\parindent}It should be remarked that any model of the
 spontaneous CP violation suffers from the cosmological domain wall
 problem.
 In the present paper we are interested in constructing simple
 examples of the dynamical CP violation and we tentatively
 circumvent the problem by assuming that the dynamical CP violation
 takes place before the inflation period.
\vglue 0.6cm
\newpage

\stepcounter{subsection}
\subsection{DASHEN MECHANISM IN COMPOSITE HIGGS MODELS}
\stepcounter{section}
\vglue 0.1cm
\hspace*{\parindent}%
Here in the present section we briefly review
 the Dashen mechanism of the spontaneous CP violation with the
 application to the composite Higgs models.\\
\hspace*{\parindent}We start with the Lagrangian ${\cal L}_0$
 symmetric under the flavor group
\begin{equation}
       G_{F}=\prod_{\rho}U_{V}(n_{\rho})\otimes U_{A}(n_{\rho})\ ,
\end{equation}
\noindent
where $n_{\rho}$ is the number of quark flavors belonging to the
 irreducible representation $\rho$ in the underlying gauge group
 and $U_V$ ($U_A$) is the unitary group associated with
 the vector (axialvector) currents.
 Here by the term ``quark'' we mean the ordinary quarks as well as
 the fermions needed to generate the composite Higgs field.
 The quark fields included in the Lagrangian ${\cal L}_0$ are all
 massless to guarantee the underlying gauge symmetry and the
 flavor symmetry.\\
\hspace*{\parindent}We assume that the flavor symmetry $G_F$ is
 broken dynamically by the presence of the nonvanishing vacuum
 expectation value for the composite field made of certain
quark fields,
\begin{equation}
       \langle \bar{\psi}\psi \rangle \neq 0\ .
\end{equation}
Here we have chosen the vacuum for which
\begin{equation}
       \langle \bar{\psi}\gamma_{5}\psi \rangle = 0\ .
\end{equation}
The quarks acquire masses according to the dynamical breaking of
 the flavor symmetry $G_F$.
 The remaining flavor symmetry if any will be denoted by $S_F$.
 As is wellknown the vacuum satisfying Eqs.$(2.2)$ and $(2.3)$ is
 not unique and thus we have degenerate vacua in $G_{F}/S_{F}$.
These degenerate vacua point to arbitrary direction in
 $G_{F}/S_{F}$.\\
\hspace*{\parindent}Now we add to ${\cal L}_0$ the term $\cal L'$
 which explicitly breaks the flavor symmetry $G_F$.
 We assume that $\cal L'$ is CP-invariant and $S_F$-symmetric.
 The degeneracy of the vacua mentioned above is now resolved in
 the system described by the total Lagrangian
\begin{equation}
       {\cal L}={\cal L}_0+{\cal L'}\ .
\end{equation}
The direction of the vacuum thus determined, however, does not
 necessarily guarantee the conditions $(2.2)$ and $(2.3)$.
 Hence we need to make a transformation on the field to recover
 the conditions
\begin{equation}
       {\psi}'=U\psi\ ,
\end{equation}
\noindent
with $U$ the transformation belonging to $G_F$.
 By this transformation the form of the symmetry breaking term
 $\cal L'$ will be modified so that CP violating terms in general
 show up in $\cal L'$.
 We will call this mechanism of the spontaneous CP
 violation\cite{ODCP} the Dashen mechanism.
 In the following we would like to apply the Dashen mechanism to
 the case of the composite Higgs models.\\
\hspace*{\parindent}In electroweak theory the Higgs fields are
 introduced as elementary scalar fields.
 Accordingly the Higgs mass, Higgs self-coupling constant and
 Higgs-fermion Yukawa-coupling constants are all arbitrary
 parameters.
 In the composite Higgs model the Higgs particle appears as a
 composite system of some fundamental fermions and some of the
 parameters in the standard electroweak theory are predictable in
 principle.
 The Lagrangian corresponding to this model may be given by
\begin{equation}
       {\cal L}_0={\cal L}_{QCD}+{\cal L}_{EW}+{\cal L}_{DYN}\ ,
\end{equation}
\noindent
where ${\cal L}_{QCD}$ is the ordinary QCD Lagrangian for quarks,
 ${\cal L}_{EW}$ is the electroweak Lagrangian without Higgs fields
 and ${\cal L}_{DYN}$ is the dynamical term which is assumed to be
 responsible for generating the composite Higgs system as a bound
 state (this term may be thought of as a low-energy effective
 Lagrangian stemming from the more fundamental Lagrangian).\\
\hspace*{\parindent}The Higgs particle appears as a bound state
 of the fundamental fermions $\psi$ and the bound state is assumed
 to generate a condensation,
\begin{equation}
       \langle \bar{\psi}\psi \rangle \neq 0\ .
\end{equation}
\noindent
The fundamental fermions as well as the ordinary quarks acquire
 a mass according to the condensation.
 The mass of the fundamental fermions should be of the order of
 the weak scale in order to guarantee that the resulting effective
 theory be the standard electroweak theory.\\
\hspace*{\parindent}In the technicolor model\cite{TECH} the
 fundamental fermion is the techniquark,
 in the top-condensation model\cite{TOPC} it is the top quark
 with mass close to the weak scale,
 in the fourth-generation model\cite{FG} it is the heavy quark in
 the assumed fourth generation and in the color-sextet model\cite{SQ}
 it is the quark in the sextet representation of the color SU(3).\\
\hspace*{\parindent}In the following we would like to present
 simple models of the dynamical CP violation in the composite Higgs
 models.
\vglue 0.6cm

\setcounter{equation}{0}
\stepcounter{subsection}
\stepcounter{subsection}
\subsection{SIMPLE MODELS OF DYNAMICAL CP VIOLATION}
\stepcounter{section}
{\elevenbf\noindent A. General formalism}
\vglue 0.3cm
Here we first present a general argument in constructing simple
 models of the dynamical CP violation in the composite Higgs models.
 We consider $n_{\rho}$ flavors of fundamental quarks in the
 representation $\rho$ of the color SU(3) or other symmetry group
 (we call this symmetry governing the fundamental quarks the
 symmetry S)
 and $n_3$ flavors of ordinary quarks in the triplet representation
 of the color SU(3).
 The fundamental quarks may or may not have a color degree of
 freedom.\\
\hspace*{\parindent}We will discuss transformations which mix the
 flavors of the fundamental and ordinary quarks among themselves.
 Since this transformation has to conserve charges, the mixing
 occurs only among the up-type (or down-type) fundamental and
 ordinary quarks.
 For simplicity we consider only up-type fundamental and ordinary
 quarks.\\
\hspace*{\parindent}According to Goldstein's analysis\cite{DCP2}
 one finds that only two flavors of the fundamental and ordinary
 quarks are not sufficient to realize the Dashen mechanism.
 Hence we try a model with 2 flavors of the up-type fundamental
 quarks $Q$ and 3 flavors of the up-type ordinary quarks $q$:
\begin{equation}
       Q=(U,C)\ ,\mbox{\hspace{1cm}} \q=(u,c,t)\ .
\end{equation}
\noindent
We assume that $Q$ belongs to the N-plet of the fundamental
 symmetry S and $q$ belongs to the color triplet.
 It is understood that our model equally applies to the system
 of the down-type quarks
\begin{equation}
       Q=(D,S)\ ,\mbox{\hspace{0.9cm}}\q=(d,s,b)\ .
\end{equation}
\noindent
 In the following by the term ``quark'' we generically mean both
 fundamental and ordinary quarks.\\
\hspace*{\parindent}As a $G_F$ breaking Hamiltonian density
 $\cal H'$ we take the following four-fermion terms
\begin{eqnarray}
       {\cal H'}=-{\cal L'}=\!\!\!
        &   & \!\!\!{G^{Q}}_{\alpha \beta}
               \bar{Q}_{\sL}\tau^{\alpha}Q_{\sR}\bar{Q}_{\sR}
               \tau^{\beta}Q_{\sL}                  \nonumber \\
 \!\!\! & + & \!\!\!{G^{Qq}}_{\alpha \beta}
               \bar{Q}_{\sL}\tau^{\alpha}Q_{\sR}\bar{\q}_{\sR}
               \lambda^{\beta}\q_{\sL}
               +h.c.                                \nonumber \\
 \!\!\! & + & \!\!\!{G^{q}}_{\alpha \beta}
               \bar{\q}_{\sL}\lambda^{\alpha}\q_{\sR}\bar{\q}_{\sR}
               \lambda^{\beta}\q_{\sL}\ ,
\end{eqnarray}
\newpage
\noindent
where ${G^{Q}}_{\alpha \beta}$ , ${G^{Qq}}_{\alpha \beta}$
 and ${G^{q}}_{\alpha \beta}$ are coupling parameters among
 fundamental quarks $Q$ and ordinary quarks $\q$ which depend
 on indices $\alpha$ and $\beta$ of the flavor SU(2) and SU(3)
 matrices $\tau^{\alpha}$ $(\alpha =1,2,3)$ and $\lambda^{\alpha}$
 $(\alpha =1,2,\ldots ,8)$ respectively.
 In Eq.$(3.3)$ the fundamental symmetry indices and color indices
 are suppressed and are understood to be contracted
 between adjoining quarks.
 There would be other possibilities of contracting these indices.
 We, however, confine ourselves to the case of Eq.$(3.3)$.\\
\hspace*{\parindent}We require the CP invariance and hermiticity
 of the Lagrangian $(3.3)$.
 We then have
\begin{eqnarray}
       &   & {G^{Q}}_{\alpha \beta}{\tau^{\alpha}}_{rr'}
             {\tau^{\beta}}_{ss'}
         =   {G^{Q}}_{\beta \alpha}{\tau^{\alpha}}_{r'r}
             {\tau^{\beta}}_{s's}
         =   ({G^{Q}}_{\beta \alpha}{\tau^{\alpha}}_{r'r}
             {\tau^{\beta}}_{s's})^{\ast}    \ ,
                                                    \nonumber \\
       &   & {G^{Qq}}_{\alpha \beta}{\tau^{\alpha}}_{rr'}
             {\lambda^{\beta}}_{ss'}
         =   ({G^{Qq}}_{\alpha \beta}{\tau^{\alpha}}_{rr'}
             {\lambda^{\beta}}_{ss'})^{\ast}\ ,
                                                    \nonumber \\
       &   & {G^{q}}_{\alpha \beta}{\lambda^{\alpha}}_{rr'}
             {\lambda^{\beta}}_{ss'}
         =   {G^{q}}_{\beta \alpha}{\lambda^{\alpha}}_{r'r}
             {\lambda^{\beta}}_{s's}
         =   ({G^{q}}_{\beta \alpha}{\lambda^{\alpha}}_{r'r}
             {\lambda^{\beta}}_{s's})^{\ast}\ ,
\end{eqnarray}
\noindent
where indices $r, r', s, s'$ represent flavors of $Q$ and $\q$,
 i.e. $U, C, u, c, t.$\\
\hspace*{\parindent}Our first task is to find the correct vacuum
 under the Lagrangian
\begin{equation}
       {\cal L}={\cal L}_0+{\cal L'}\ ,
\end{equation}
\noindent
where ${\cal L}_0$ is the Lagrangian given by Eq.$(2.6)$ and
 ${\cal L'}$ is given by Eq.$(3.3)$.
 Let us denote by $|\bar{0}\rangle$ the ground state (vacuum) for
 a system governed by the Lagrangian $(3.5)$ and by $|0\rangle$
 the ground state for ${\cal L}_0$ which is invariant under $S_F$.
 To find the ground state $|\bar{0}\rangle$ we try to minimize the
 energy
\begin{equation}
       E(W)=\langle \bar{0}\, |\, {\cal H'}\, |\, \bar{0}\rangle
           =\langle 0\, |\, W^{\scriptscriptstyle \dag}{\cal H'}W\,
            |\, 0\rangle\ ,
\end{equation}
\noindent
by suitably choosing the transformation $W$ in $G_F$.
 The transformation $W$ is induced by the transformation $U$ of
 fermion fields $Q$ and $\q$:
\begin{equation}
       Q'_{\sL,\sR}={U^{Q}}_{\sL,\sR}Q_{\sL,\sR}\ ,
       \mbox{\hspace{1cm}} \q'_{\sL,\sR}={U^{q}}_{\sL,\sR}
       \q_{\sL,\sR}\ ,
\end{equation}
\noindent
where $U^{Q}_{\sL,\sR}$ is the transformation belonging to the
 left-handed(right-handed) flavor SU(2) for fundamental quarks
 $Q$ and $U^{q}_{\sL,\sR}$ belonging to the SU(3) for ordinary
 quarks $q$.
 The transformation $W$ is a function of these fermion
 transformations:
\begin{equation}
       W=W(U)\ ,
\end{equation}
\noindent
where we represent generically by $U$ the transformations
 ${U^{Q}}_{\sL,\sR}$ and ${U^{q}}_{\sL,\sR}$.
 We find
\newpage
\begin{eqnarray}
       E(W)= \!\! & & \!\!\!\!\!\!\!\!{G^{Q}}_{\alpha \beta}
            \langle 0\, |\, \bar{Q}_{\sL}{U^{Q}}_{\sL}^{\sdag}
                      \tau^{\alpha}{U^{Q}}_{\sR}Q_{\sR}
                      \bar{Q}_{\sR}{U^{Q}}_{\sR}^{\sdag}
                      \tau^{\beta}{U^{Q}}_{\sL}Q_{\sL}\,
            |\, 0\rangle                           \nonumber \\
  \!\!  & + & \!\!\!{G^{Qq}}_{\alpha \beta}
            \langle 0\, |\, \bar{Q}_{\sL}{U^{Q}}_{\sL}^{\sdag}
                      \tau^{\alpha}{U^{Q}}_{\sR}Q_{\sR}
                      \bar{\q}_{\sR}{U^{q}}_{\sR}^{\sdag}
                      \lambda^{\beta}{U^{q}}_{\sL}\q_{\sL}
            +h.c.\, |\, 0\rangle                    \nonumber \\
  \!\!  & + & \!\!\!{G^{q}}_{\alpha \beta}
            \langle 0\, |\, \bar{\q}_{\sL}{U^{q}}_{\sL}^{\sdag}
                      \lambda^{\alpha}{U^{q}}_{\sR}\q_{\sR}
                      \bar{\q}_{\sR}{U^{q}}_{\sR}^{\sdag}
                      \lambda^{\beta}{U^{q}}_{\sL}\q_{\sL}\,
            |\, 0\rangle\ .
\end{eqnarray}
\hspace*{\parindent}Since the state $|0\rangle$ is invariant under
 $S_F$, we may express the following amplitudes as given below:
\begin{eqnarray}
       & & \langle 0\, |\, \bar{Q}_{\sL r}Q_{\sR r'}\, |\, 0\rangle
           ={\Delta^Q}_{0}\delta_{rr'}\ , \;
           \langle 0\, |\, \bar{\q}_{\sL r}\q_{\sR r'}\, |\, 0\rangle
           ={\Delta^q}_{0}\delta_{rr'}\ ,
                                                        \nonumber \\
       & & \langle 0\, |\, \bar{Q}_{\sL r}Q_{\sR r'}
           \bar{Q}_{\sR s}Q_{\sL s'}\, |\, 0\rangle =
           {\Delta^Q}\delta_{rr'}\delta_{ss'}+{\Delta^{\prime Q}}
           \delta_{rs'}\delta{r's}\ ,
                                                        \nonumber \\
       & & \langle 0\, |\, \bar{Q}_{\sL r}Q_{\sR r'}
           \bar{\q}_{\sR s}\q_{\sL s'}\, |\, 0\rangle =
           {\Delta^{Qq}}\delta_{rr'}\delta_{ss'},       \nonumber \\
       & & \langle 0\, |\, \bar{\q}_{\sL r}\q_{\sR r'}
           \bar{\q}_{\sR s}\q_{\sL s'}\, |\, 0\rangle =
           {\Delta^q}\delta_{rr'}\delta_{ss'}+{\Delta^{\prime q}}
           \delta_{rs'}\delta{r's}\ ,
\end{eqnarray}
\noindent
where parameters $\Delta$ are chosen to be real.
 After some algebra we obtain
\begin{eqnarray}
       E(W)=    \!\!
          &   & \!\!\!\!\!\!\!\!{g^{Q}}_{\alpha\beta}
                \mbox{Tr}[\, U^{Q}\tau^{\alpha}]
                \mbox{Tr}[\, \tau^{\beta}
                U^{Q\scriptscriptstyle \dag}] \nonumber \\
    \!\!  & + & \!\!\!r{g^{Qq}}_{\alpha\beta}
                \mbox{Tr}[\, U^{Q}\tau^{\alpha}]
                \mbox{Tr}[\, \lambda^{\beta}
                U^{q\scriptscriptstyle \dag}]+h.c. \nonumber \\
    \!\!  & + & \!\!\!r^2{g^{q}}_{\alpha\beta}
                \mbox{Tr}[\, U^{q}\lambda^{\alpha}]
                \mbox{Tr}[\, \lambda^{\beta}
                U^{q\scriptscriptstyle \dag}]\ ,
\end{eqnarray}
\noindent
where matrices $U^Q$ and $U^q$ and parameters
 ${g^{Q}}_{\alpha\beta}$ , ${g^{Qq}}_{\alpha\beta}$ ,
 ${g^{q}}_{\alpha\beta}$ and $r$ are given by the following
 relations:
\begin{equation}
       U^{Q}={U^Q}_{\sR}{U^Q}_{\sL}^{\scriptscriptstyle \dag}\ ,
             \mbox{\hspace{1cm}}
       U^{q}={U^q}_{\sR}{U^q}_{\sL}^{\scriptscriptstyle \dag}\ ,
\end{equation}
\begin{eqnarray}
            {g^{Q}}_{\alpha\beta}={G^{Q}}_{\alpha\beta}
            \Delta^{Q}\ ,\;
\!\!\!\! & r & \!\!\!\!\!{g^{Qq}}_{\alpha\beta}=
            {G^{Qq}}_{\alpha\beta}\Delta^{Qq}\ ,\;
         r^2{g^{q}}_{\alpha\beta}={G^{q}}_{\alpha\beta}
            \Delta^{q}\ ,\\
         & r & \!\!\!=\frac{\langle\, \bar{\q}\q\, \rangle}
            {\langle \bar{Q}Q\rangle}
                   \ .
\end{eqnarray}
Here we introduced parameter $r$ in order to show explicitly the
 relative size of the three kinds of parameters
 ${G^{Q}}_{\alpha \beta}$ , ${G^{Qq}}_{\alpha \beta}$ and
 ${G^{q}}_{\alpha \beta}$.
 The parameter $r$ is the ratio of the ordinary and fundamental
 mass scale\cite{TECH} and its size is assumed to be
\begin{equation}
       r \sim \left(\frac{1\, \mbox{GeV}}{1\, \mbox{TeV}}\right)^3
           =10^{-9}\ .
\end{equation}
Our task is to minimize $E(W)$ given in Eq.$(3.11)$ by changing $U$
 and find the solution for $U$.
 With $U$ determined in this procedure we rewrite ${\cal L'}$ to
 see whether CP violating terms are generated in ${\cal L'}$.
\newpage
{\elevenbf\noindent B. Special solutions}
\vglue 0.3cm
We would like to find the general solution for $U$ to minimize
 $E(W)$ in Eq.$(3.11)$.
 It is, however, quite complicated to obtain the general solution
 and we shall confine ourselves to some special solutions to this
 problem.\\
\hspace*{\parindent}We first consider the following specialization,
\begin{eqnarray}
       & & {g^Q}_{00}={g^Q}_{33}\; (\equiv g^Q)<0\ ,\nonumber \\
       & & 3{g^{Qq}}_{00}=-\sqrt{3}\ {g^{Qq}}_{08}
           =\sqrt{3}\ {g^{Qq}}_{30}=2{g^{Qq}}_{38}
                             \; (\equiv g^{Qq})>0\ ,\nonumber \\
       & & {g^{Q}}_{\alpha \beta}=0,\; {g^{Qq}}_{\alpha \beta}=
           0\; : \mbox{ otherwise}\ .
\end{eqnarray}
\noindent
Parametrizing the matrix elements of matrices $U^Q$ and $U^q$ by
\begin{eqnarray}
       & & {U^{Q}}_{i j}={u^{Q}}_{i j}\exp (i{\theta^{Q}}_{ij})\;
                  \mbox{\hspace{2ex}with }\; i,j=1,2\nonumber \\
       & & {U^{q}}_{i j}={u^{q}}_{i j}\exp (i{\theta^{q}}_{ij})\;
                  \mbox{\hspace{4ex}with }\; i,j=1,2,3
\end{eqnarray}
\noindent
where $u$'s and $\theta$'s are real constants constrained by the
 unitality of $U$, we obtain
\begin{eqnarray}
       E(W) \!\! & = & \!\!2g^Q\{({u^{Q}}_{11})^2+({u^{Q}}_{22})^2\}
                                                        \nonumber \\
            &   & \!\!\!\!+2rg^{Qq}\left\{\frac{\sqrt{3}}{2}
                  {u^Q}_{11}{u^q}_{11}
                  \cos({\theta^Q}_{11}-{\theta^q}_{11})
                                   \right.              \nonumber \\
            &   & \mbox{\hspace{5ex}}
                  +\frac{\sqrt{3}}{2}{u^Q}_{11}{u^q}_{22}
                  \cos({\theta^Q}_{11}-{\theta^q}_{22})
                                                        \nonumber \\
            &   & \mbox{\hspace{5ex}}
                  +{u^Q}_{11}{u^q}_{33}
                  \cos({\theta^Q}_{11}-{\theta^q}_{33})
                                                        \nonumber \\
            &   & \mbox{\hspace{5ex}}
                  -\frac{\sqrt{3}}{2}{u^Q}_{22}{u^q}_{11}
                        \cos({\theta^Q}_{22}-{\theta^q}_{11})
                                                        \nonumber \\
            &   & \mbox{\hspace{5ex}}
                  -\frac{\sqrt{3}}{2}{u^Q}_{22}{u^q}_{22}
                  \cos({\theta^Q}_{22}-{\theta^q}_{22})
                                                        \nonumber \\
            &   & \mbox{\hspace{5ex}}
                  +{u^Q}_{22}{u^q}_{33}
                  \cos({\theta^Q}_{22}-{\theta^q}_{33})
                                                   \Biggr\}
                                                        \nonumber \\
            &   & \!\!\!\!+\mbox{O}(r^2)\ .
\end{eqnarray}
\noindent
In deriving Eq.$(3.18)$ we kept only the terms up to the first
 order of the small number $r$.
 We expand parameters $u$'s and $\theta$'s in powers of $r$ and
 look for the minimum of the energy $E(W)$ to the first order
 of $r$:
\begin{eqnarray}
       & & E=E^0+E^1r+\mbox{O}(r^2)\ ,\nonumber \\
       & & u=u^0+u^1r+\mbox{O}(r^2)\ ,\nonumber \\
       & & \theta =\theta^0+\theta^1r+\mbox{O}(r^2)\ ,
\end{eqnarray}
\noindent
where we have omitted the suffices $i$ and $j$ and the superfix
 $Q$ or $q$ in the parameters $u$ and $\theta$.
 After some algebra we find
\begin{equation}
       E^0 = 2g^Q\{({u^{Q0}}_{11})^2+({u^{Q0}}_{22})^2\}\ ,
\end{equation}
\begin{eqnarray}
       E^1 & = & \!\!4g^Q({u^{Q0}}_{11}{u^{Q1}}_{11}
                +{u^{Q0}}_{22}{u^{Q1}}_{22})
                                                       \nonumber \\
           &   & \!\!\!\!+2g^{Qq}\left\{\frac{\sqrt{3}}{2}
                 {u^{Q0}}_{11}{u^{q0}}_{11}
                 \cos({\theta^{Q0}}_{11}-{\theta^{q0}}_{11})
                                   \right.             \nonumber \\
           &   & \mbox{\hspace{5ex}}
                 +\frac{\sqrt{3}}{2}{u^{Q0}}_{11}{u^{q0}}_{22}
                 \cos({\theta^{Q0}}_{11}-{\theta^{q0}}_{22})
                                                       \nonumber \\
           &   & \mbox{\hspace{5ex}}
                 +{u^{Q0}}_{11}{u^{q0}}_{33}
                 \cos({\theta^{Q0}}_{11}-{\theta^{q0}}_{33})
                                                       \nonumber \\
           &   & \mbox{\hspace{5ex}}
                 -\frac{\sqrt{3}}{2}{u^{Q0}}_{22}{u^{q0}}_{11}
                 \cos({\theta^{Q0}}_{22}-{\theta^{q0}}_{11})
                                                       \nonumber \\
           &   & \mbox{\hspace{5ex}}
                 -\frac{\sqrt{3}}{2}{u^{Q0}}_{22}{u^{q0}}_{22}
                 \cos({\theta^{Q0}}_{22}-{\theta^{q0}}_{22})
                                                       \nonumber \\
           &   & \mbox{\hspace{5ex}}
                 +{u^{Q0}}_{22}{u^{q0}}_{33}
                 \cos({\theta^{Q0}}_{22}-{\theta^{q0}}_{33})
                                                  \Biggr\}\ .
\end{eqnarray}
\noindent
Since we chose $g^Q<0$, ${u^{Q0}}_{11}$ and ${u^{Q0}}_{22}$ may
 be taken to be unity according to Eq.$(3.20)$.
 We find the following set of parameters $u$'s and $\theta$'s to
 minimize $E^1$.
\begin{eqnarray}
       & & {u^{Q0}}_{11}={u^{Q0}}_{22}=1\ ,             \nonumber \\
       & & {u^{Q1}}_{11}={u^{Q1}}_{22}=0\ ,             \nonumber \\
       & & {u^{q0}}_{11}={u^{q0}}_{22}={u^{q0}}_{33}=1\ ,
                                                        \nonumber \\
       & & {\theta^{Q0}}_{11}=\theta\pm \frac{\pi}{3}\ ,\;
           {\theta^{Q0}}_{22}=\theta\mp \frac{\pi}{3}\ ,\;
                        \mbox{\hspace{13ex}}\pmod{2\pi} \nonumber \\
       & & {\theta^{q0}}_{11}=\theta\mp \frac{\pi}{2}\ ,\;
           {\theta^{q0}}_{22}=\theta\mp \frac{\pi}{2}\ ,\;
           {\theta^{q0}}_{33}=\theta\pm \pi          \ ,\;
                        \mbox{\hspace{1ex}}\pmod{2\pi}
\end{eqnarray}
\noindent
where $\theta$ is the free parameter.
 Thus the transformation matrices $U^Q$ and $U^q$ are given by
\begin{eqnarray}
       U^Q & \!\!\!\! = {U^Q}_{\sR}{U^Q}_{\sL}^{\sdag}
                      = & \!\!\! e^{i\theta}
       \left(
             \begin{array}{cc}
              e^{\pm i\frac{\pi}{3}} & 0                        \\
                        0            & e^{\mp i\frac{\pi}{3}}
             \end{array}
       \right)\ ,                                     \nonumber \\
       U^q & \!\!\!\! = {U^q}_{\sR}{U^q}_{\sL}^{\sdag}
                      = & \!\!\!\! e^{i\theta}
       \left(
             \begin{array}{ccc}
              e^{\mp i\frac{\pi}{2}} &           0
                                     &     0        \\
                        0            & e^{\mp i\frac{\pi}{2}}
                                     &     0        \\
                        0            &           0
                                     & e^{\pm i\pi}
             \end{array}
       \right)\ .
\end{eqnarray}
\noindent
In the present paper we would like to construct a model without
 the strong CP violation and so we set
\begin{equation}
       \theta = 0 \ .
\end{equation}
\newpage
We redefine the fields $Q$ and $\q$ in such a way that
\begin{eqnarray}
       Q'_{\sL}=Q_{\sL}\ , & \mbox{\hspace{0.7cm}} &
       Q'_{\sR}={U^Q}_{\sL}{U^Q}_{\sR}^{\sdag}Q_{\sR}\ , \nonumber \\
       q'_{\sL}=q_{\sL}\ , & \mbox{\hspace{0.7cm}} &
       q'_{\sR}={U^q}_{\sL}{U^q}_{\sR}^{\sdag}q_{\sR}\ .
\end{eqnarray}
\noindent
so that the following condition is recovered,
\begin{eqnarray}
       & & \langle 0\, |\, \bar{Q'}_{\sL r}
                        Q'_{\sR r'}\, |\, 0\rangle
           ={\Delta^Q}_{0}\delta_{rr'}\ ,               \nonumber \\
       & & \langle 0\, |\, \bar{\q'}_{\sL r}
                        \q'_{\sR r'}\, |\, 0\rangle
           ={\Delta^q}_{0}\delta_{rr'}\ .
\end{eqnarray}
\noindent
Expressed by the new fields $Q'$ and $\q'$ the Hamiltonian
 density $\cal H'$ takes the form
\begin{eqnarray}
      {\cal H'}=\!\!\!
        &   & \!\!\!\!\!\!-{\cal L'}={G^{Q}}_{\alpha \beta}
               \bar{Q'}_{\sL}\tau^{\alpha}{U^Q}_{\sR}
              {U^Q}_{\sL}^{\sdag}Q'_{\sR}
               \bar{Q'}_{\sR}{U^Q}_{\sL}{U^Q}_{\sR}^{\sdag}
               \tau^{\beta}Q'_{\sL}
                                                        \nonumber \\
 \!\!\! & + & \!\!\!{G^{Qq}}_{\alpha \beta}
               \bar{Q'}_{\sL}\tau^{\alpha}{U^Q}_{\sR}
               {U^Q}_{\sL}^{\sdag}Q'_{\sR}
               \bar{\q'}_{\sR}{U^q}_{\sL}{U^q}_{\sR}^{\sdag}
               \lambda^{\beta}\q'_{\sL}
               +h.c.                                    \nonumber \\
 \!\!\! & + & \!\!\!{G^{q}}_{\alpha \beta}
               \bar{\q'}_{\sL}\lambda^{\alpha}{U^q}_{\sR}
               {U^q}_{\sL}^{\sdag}\q'_{\sR}
               \bar{\q'}_{\sR}{U^q}_{\sL}{U^q}_{\sR}^{\sdag}
               \lambda^{\beta}\q'_{\sL}\ .
\end{eqnarray}
\noindent
We find in Eq.$(3.27)$ that the second and third term in general
 violate CP since these two terms can not be made real.
 To see this situation more explicitly we rewrite Eq.$(3.27)$
 in the following form,
\begin{eqnarray}
      {\cal H'}=\!\!\!
        &   & \!\!\!2{G^{Q}}
               (\bar{U'}_{\sL}U'_{\sR}\bar{U'}_{\sR}U'_{\sL}
               +\bar{C'}_{\sL}C'_{\sR}\bar{C'}_{\sR}C'_{\sL})
                                                      \nonumber \\
 \!\!\! & + & \!\!\!{G^{Qq}}\biggl[
               \frac{\sqrt{3}}{2}(e^{i\frac{5}{6}\pi}
               \bar{U'}_{\sL}U'_{\sR}
               -e^{i\frac{1}{6}\pi}\bar{C'}_{\sL}C'_{\sR})
              (\bar{u'}_{\sR}u'_{\sL}+\bar{c'}_{\sR}c'_{\sL})
                                                      \nonumber \\
        &   & \mbox{\hspace*{4ex}}+(e^{-i\frac{2}{3}\pi}
              \bar{U'}_{\sL}U'_{\sR}
              +e^{-i\frac{4}{3}\pi}\bar{C'}_{\sL}C'_{\sR})
                                 (\bar{t'}_{\sR}t'_{\sL})
                            \biggr]
               +h.c.                                  \nonumber \\
 \!\!\! & + & \!\!\!\frac{4}{3}{G^{q}}_{88}
               \bar{t'}_{\sL}t'_{\sR}\bar{t'}_{\sR}t'_{\sL}+\cdots
                                                      \nonumber \\
 \!\!\! & + & \!\!\!({G^{q}}_{44}-{G^{q}}_{55})e^{i\frac{3}{2}\pi}
               \bar{u'}_{\sL}t'_{\sR}\bar{u'}_{\sR}t'_{\sL}+\cdots\; .
\end{eqnarray}
\noindent
We now clearly observe that the new expression of the Hamiltonian
 density $\cal H'$ includes CP violating terms.
 In fact the second and fourth term in $\cal H'$ apparently
 violate CP.
 On the other hand the other terms do not violate CP.\\
\hspace*{\parindent}
It may be interesting to see the form of the up-type quark mass
 matrix $(M_{ij})$ in the present specific model.
 After some algebra we find
\begin{equation}
       (M_{ij})=2g^{Qq}
               \left(
               \begin{array}{ccc}
                   -3/4    &    0   &    0   \\
                     0     &  -3/4  &    0   \\
                     0     &    0   &  -1/2
               \end{array}
               \right)\ .
\end{equation}
\noindent
 Obviously the form $(3.29)$ does not properly reproduce the
 quark mass hierarchy and so our model is not realistic.
\newpage
We next consider the following specific choice of our coupling
 parameters,
\begin{eqnarray}
       & & {g^Q}_{00}={g^Q}_{33}\; (\equiv g^Q)<0\ ,\nonumber \\
       & & \sqrt{3}{g^{Qq}}_{00}=-\ {g^{Qq}}_{08}
           =\sqrt{3}\ {g^{Qq}}_{30}=2\ {g^{Qq}}_{38}
                             \; (\equiv g^{Qq})>0\ ,\nonumber \\
       & & {g^{Q}}_{\alpha \beta}=0,\; {g^{Qq}}_{\alpha \beta}
           =0\; : \mbox{ otherwise}\ .
\end{eqnarray}
\noindent
Following the same procedure as before we find the solution
\begin{eqnarray}
       U^Q & \!\!\!\! = {U^Q}_{\sR}{U^Q}_{\sL}^{\sdag}
         = & \!\!\! e^{i\theta}
       \left(
             \begin{array}{cc}
              e^{\pm i\frac{\pi}{4}} & 0                        \\
                        0            & e^{\mp i\frac{\pi}{4}}
             \end{array}
       \right)\ ,                                     \nonumber \\
       U^q & \!\!\!\! = {U^q}_{\sR}{U^q}_{\sL}^{\sdag}
                      = & \!\!\!\! e^{i\theta}
       \left(
             \begin{array}{ccc}
              e^{\mp i\frac{\pi}{2}} &           0
                                     &     0        \\
                        0            & e^{\mp i\frac{\pi}{2}}
                                     &     0        \\
                        0            &           0
                                     & e^{\pm i\pi}
             \end{array}
       \right)\ .
\end{eqnarray}
\noindent
The up-type quark mass matrix corresponding to the above solution
 reads
\begin{equation}
       (M_{ij})=2g^{Qq}
               \left(
               \begin{array}{ccc}
                   -\sqrt{6}/4    &    0   &    0   \\
                     0     &  -\sqrt{6}/4  &    0   \\
                     0     &    0   &  -\sqrt{6}/2
               \end{array}
               \right)\ .
\end{equation}
We observe that in this case the top quark is heavier than the up
 and charm quark.
\vglue 0.4cm

{\elevenbf\noindent C. Models}
\vglue 0.3cm
We found the CP violating interaction Lagrangian as a result of
 special solutions of the minimum $E(W)$ condition.
 Thus we succeeded in constructing the simple model of the dynamical
 CP violation.
 In deriving the model we made some simplifying assumptions.
 This simplification made the model far from explaining the real
 situation in standard theory.
 For example our model Hamiltonian does not reproduce the KM
 matrix correctly.
 In order to get the full KM matrix we have to relax our assumptions
 and minimize $E(W)$ with the full expression of the transformation
 matrix $U$
 (We have to abolish the assumption that $U$ be a diagonal matrix).
 This attempt will be made in a separate work.
 We are, however, interested in estimating physical effects in
 low-energy phenomena which are predicted by the Hamiltonian.
 Such estimation may help examining whether our model serves as
 a prototype of the real theory of the dynamical CP violation for
 standard theory.\\
\hspace*{\parindent}The system of quarks we assumed consists of
 the up-type 2-flavor fundamental quarks $Q$ and 3-flavor ordinary
 quarks $q$ as shown in Eq.$(3.1)$.
 We have not yet specified the symmetry group S to which the
 fundamental quarks $Q$ belong.\\
\hspace*{\parindent}A natural possibility is to identify the
 symmetry group S to the technicolor SU(N).
 In this case the fundamental quark $Q$ is the techniquark\cite{TECH}
 belonging to the N-dimensional fundamental representation of the
 technicolor SU(N).\\
\hspace*{\parindent}Another possibility is to identify the
 symmetry group S to the color SU(3).
 In this case the fundamental quark $Q$ is the color-sextet
 quark\cite{SQ} belonging to the 6-dimensional representation of
 the color SU(3).\\
\hspace*{\parindent}These two possibilities fit the previous
 argument quite well and constitute two practical models of
 the dynamical CP violation.\\
\hspace*{\parindent}It is also possible to identify the fundamental
 quarks $Q$ to the top quark in the top-condensation model\cite{TOPC}
 (or in the top-color model\cite{TCOL}).
 In this case, however, we are not able to get the nontrivial CP
 violating phase within our framework\\
\hspace*{\parindent}Yet another possibility is to identify the
 fundamental fermion $Q$ to the quark in the assumed fourth
 generation\cite{FG}.
 In this case again it is impossible to obtain the nontrivial CP
 violating phase in our approach.\\
\hspace*{\parindent}In the following application we are in mind the
 technicolor model as well as the color-sextet quark model.
\vglue 1cm

\setcounter{equation}{0}
\stepcounter{subsection}
\stepcounter{subsection}
\stepcounter{subsection}
\subsection{LOW ENERGY EFFECTS}
\stepcounter{section}
\hspace*{\parindent}%
In our simple model introduced in the last section the KM matrix is
 real and diagonal.
 This is because we have taken a paticular choice for a $G_F$
 breaking Lagrngian $\cal L'$ and have neglected the higher order
 terms in $r$.
 Starting with the more general assumption we could have obtained
 the KM matrix with off-diagonal elements and complex phases.\\
\hspace*{\parindent}In the present section we consider possible
 low-energy effects originating from the model Lagrangian $(3.27)$.
 By this analysis we will be able to compare the low-energy
 CP-violating effects of dynamical origin with the one in the
 standard origin of the CP violation (i.e. through the KM phase).\\
\hspace*{\parindent}Our Hamiltonian reads
\begin{equation}
       H= H_{0}+H'_{cons}+H'_{viol}\ ,
\end{equation}
\noindent
where $H_{0}$ is the Hamiltonian derived from Lagrangian
 ${\cal L}_{0}$,
 $H'_{cons}$ is the CP conserving part of the Hamiltonian
 defined by integrating Eq.$(3.27)$ over the space variables
 and $H'_{viol}$ is the CP violating part.
 In the following we consider two typical low-energy effects
 derived from the Hamiltonian $(4.1)$.
\vglue 0.4cm

{\elevenbf\noindent A. Neutron electric dipole moment}
\vglue 0.3cm
Since the Lagrangian $\cal L'$ includes the energy scale at
 which the four-fermion interactions are induced from the more
 fundamental gauge theory,
 it is expected that our estimate of the neutron electric dipole
 moment depend on this energy scale.
 This means that this fundamental energy scale, i.e. the cut-off
 parameter $\Lambda$,
 may be constrained by the experimental information on the
 neutron electric dipole moment.\\
\hspace*{\parindent}We estimate the size of the contribution to
 the neutron electric dipole moment coming from our CP-violating
 Lagrangian $\cal L'$ given in Eq.$(3.27)$.\\
\hspace*{\parindent}The neutron electric dipole moment $d_n$ is
 given in terms of the quark dipole moments $d_u$ and $d_d$ in
 the naive quark model such that
\begin{equation}
       d_n=\frac{4d_d-d_u}{3}\ .
\end{equation}
\noindent
The electric dipole moment of quarks is calculated through the
 following term in the quark electromagnetic form factor at zero
 momentum transfer,
\begin{equation}
       -d_q\bar{u}\sigma_{\mu \nu}\gamma_{5}q^{\nu}u\ ,
\end{equation}
\noindent
where suffix $\q$ represents the u or d quark and $\q^{\nu}$ is
 the momentum transfer for quarks (momentum carried by the virtual
 photon) and $u$ is the Dirac spinor for quark $\q$.\\
\hspace*{\parindent}We start with the Lagrangian $\cal L'$ given
 in Eq.$(3.27)$.
 For the later calculational convenience we introduce auxiliary
 fields $\phi$ and use the following effective Lagrangian instead
 of the four-fermion type Lagrangian $(3.27)$:
\begin{equation}
      {\cal L'}=-\bar{\psi}
                 \left(
                     \phi^{\sdag}\frac{1-\gamma_5}{2}
                    +\phi\frac{1+\gamma_5}{2}
                 \right)
                 \psi
                +{G^{-1}}\phi^{\sdag}\phi\ .
\end{equation}
\noindent
The use of the above auxiliary-field Lagrangian makes it easier
 to classify the relevant Feynman diagrams contributing to the
 quark electric dipole moment and to perform the higher-order loop
 calculations.\\
\hspace*{\parindent}At one-loop level diagrams shown in Fig.1
 contribute to the quark electromagnetic form factor.
\newpage
\vspace*{2cm}
\begin{center}
Fig.1a\hspace{7cm}Fig.1b
\vglue 0.3cm
{ \tenrm\baselineskip=12pt
 \noindent
Fig.1 One-loop diagrams for the electromagnetic vertex function
 of quarks\\
 represented by the use of auxiliary field $\phi$.}
\vglue 1.5cm
\end{center}
\noindent
 As is easily seen the diagram in Fig.1a has no tensor structure
 corresponding to the electric dipole moment.
 The contribution of Fig.1b to the electric dipole moment is found
 to vanish.
 Thus there is no one-loop contribution to the quark electric dipole
 moment.\\
\hspace*{\parindent}We next examine the two-loop contribution to
 the quark electric dipole moment.
 The relevant diagrams are shown in Fig.2.
\vspace{2cm}
\begin{center}
Fig.2a\hspace{7cm}Fig.2b
\end{center}
\vspace{2cm}
\begin{center}
Fig.2c
\vglue 0.3cm
{\rightskip=3pc
 \leftskip=3pc
 \tenrm\baselineskip=12pt
 \noindent
Fig.2 Two-loop diagrams for the electromagnetic vertex function
 of quarks.}
\vglue 1.5cm
\end{center}
The Feynman amplitudes corresponding to these diagrams are in
 general quartically divergent.
 The quartically divergent part of the amplitudes, however,
 has no tensor structure of the electric dipole moment and hence
 the leading contribution of these diagrams to the quark electric
 dipole moment is quadratically divergent.
 As is seen by direct calculations, the diagrams in Figs.2a and 2b
 have no quadratically divergent contribution to the quark electric
 dipole moment.
 The reason for this is that the helicity of the quark flips three
 times in these diagrams.
 Accordingly the leading quadratic divergence exists only in the
 diagram in Fig.2c.
 In the following we will calculate the quadratically divergent
 part of the Feynman amplitude corresponding to the diagram
 in Fig.2c.\\
\hspace*{\parindent}The Feynman amplitude $F$ corresponding to the
 diagram in Fig.2c reads
\begin{eqnarray}
      F & = & \sum_{i,j,k}\int\frac{d^4\!p}{(2\pi)^4i}
              \frac{d^4\!p'}{(2\pi)^4i}
                        \biggl[G^q_{jiuk}G^q_{kjiu}
                                                       \nonumber \\
        &   & \mbox{\hspace{19ex}}\times\biggl\{\frac{1+\gamma_5}{2}
                        \frac{1}{m_i-\ps_1+\ps'}
                        \frac{1-\gamma_5}{2}
                        \frac{1}{m_j-\ps_1+\ps+\ps'}
                                                       \nonumber \\
        &   & \mbox{\hspace{21ex}}\times\frac{1-\gamma_5}{2}
                        \frac{1}{m_k-\ps_1+\ps}
                        (Q_ke\gamma_{\mu})
                        \frac{1}{m_k-\ps_2+\ps}
                        \frac{1+\gamma_5}{2}
                                                        \nonumber \\
        &   & \mbox{\hspace{21ex}}+\frac{1+\gamma_5}{2}
                        \frac{1}{m_i-\ps_1+\ps'}
                        \frac{1-\gamma_5}{2}
                        \frac{1}{m_j-\ps_1+\ps+\ps'}
                        (Q_je\gamma_{\mu})
                                                        \nonumber \\
        &   & \mbox{\hspace{21ex}}\times\frac{1}{m_j-\ps_2+\ps+\ps'}
                        \frac{1-\gamma_5}{2}
                        \frac{1}{m_k-\ps_2+\ps}
                        \frac{1+\gamma_5}{2}
                                                        \nonumber \\
        &   & \mbox{\hspace{21ex}}+\frac{1+\gamma_5}{2}
                        \frac{1}{m_i-\ps_1+\ps'}
                        (Q_ie\gamma_{\mu})
                        \frac{1}{m_j-\ps_2+\ps'}
                        \frac{1-\gamma_5}{2}
                                                        \nonumber \\
        &   & \mbox{\hspace{21ex}}\times\frac{1}{m_j-\ps_2+\ps+\ps'}
                        \frac{1-\gamma_5}{2}
                        \frac{1}{m_k-\ps_2+\ps}
                        \frac{1+\gamma_5}{2}\biggr\}
                                                        \nonumber \\
        &   & \mbox{\hspace{17ex}}+G^q_{ukji}G^q_{iukj}
              \biggl\{\gamma_5 \rightarrow -\gamma_5\biggr\}
                        \biggr]\ ,
\end{eqnarray}
\noindent
where $p_1$ $(p_2)$ is the momentum of the incoming (outgoing)
 quark.
Here in Eq.$(4.5)$ the charge $Q_j$ is equal to $2/3$ for up-type
 quarks, i.e. $j = u, c, t,$ and is equal to $-1/3$ for down-type
 quarks, i.e. $j = d, s, b$.
 By extracting the quadratically divergent part $F^{div}$ of
 Eq.$(4.5)$ we obtain
\begin{equation}
      F^{div}=\frac{2\Lambda^2}{(4\pi)^4}\sum_{i,j,k}Qe
              \mbox{Im}\{G^q_{jiuk}G^q_{kjiu}\}
               m_j(iA_jk_{\mu}\gamma_5
               -B_j\sigma_{\mu\nu}\q^{\nu}\gamma_5)\ ,
\end{equation}
\noindent
where $A_j$ and $B_j$ are given by
\begin{eqnarray}
       A_j & = & 2\left(\ln\frac{\Lambda^2}{m_j^2}-1\right)
                                                       \nonumber \\
           & + & \!\!\!\!\int^{1}_{0}d\!x\int^{1}_{0}d\!y
                                         \int^{1-y}_{0}d\!z
                  \biggl[\left\{\frac{4}{x}(2-3y)+(3-5y)\right\}
                          \ln\left|\frac{x(1-x)+1-y-z}{1-y-z}\right|
                                                       \nonumber \\
           &   & \mbox{\hspace{18ex}}
                 -\frac{3x(1-x)(1-2y)}{x(1-x)+1-y-z}
                                                       \nonumber \\
           &   & \mbox{\hspace{18ex}}
                 +\frac{1}{2}\frac{x(1-x)}{\{x(1-x)+1-y-z\}^2}
                  \biggl\{x(1-x)\left(3(3-7y)-\frac{8}{x}(2-3y)\right)
                                                       \nonumber \\
           &   & \mbox{\hspace{38ex}}
                 +(1-y-z)\left(2(3-7y)
                 -\frac{6}{x}(2-3y)\right)\biggr\}\biggl]\ ,
                                                       \nonumber \\
       B_j & = & \int^{1}_{0}d\!x\int^{1}_{0}d\!y
                                 \int^{1-y}_{0}d\!z\frac{1}{x}
                 \biggl(4\ln\left|\frac{x(1-x)+1-y-z}{1-y-z}\right|
                                                       \nonumber \\
           &   & \mbox{\hspace{22ex}}
                 -\frac{x(1-x)\{4x(1-x)+3(1-y-z)\}}
                 {\{x(1-x)+1-y-z\}^2}\biggr)\ .
\end{eqnarray}
\noindent
After some algebra we derive the following formula for the
 quadratically divergent part of the electric dipole moment of the
 up-quark $d_u$
\begin{equation}
       d_u=\frac{2}{3}e\frac{\Lambda^2}{(4\pi)^4}\sum_{i,j,k}
           \mbox{Im}\{G^q_{jiuk}G^q_{kjiu}\}m_j(A_j+B_j)\ .
\end{equation}
Performing the integration in Eq.$(4.7)$ we finally find the
 explicit expression for the quadratically divergent part of
 the up-quark electric dipole moment,
\begin{equation}
       d_u=\frac{2}{3}e\frac{2\Lambda^2}{(4\pi)^4}\sum_{i,j,k}
           \mbox{Im}\{G^q_{jiuk}G^q_{kjiu}\}m_j
           \left[2\ln \frac{\Lambda^2}{m_j^2}
                                                    -2.57\right]\ .
\end{equation}
\hspace*{\parindent}Apparently the dominant contribution in the
 above formula to the up-quark dipole moment comes from the
 top-quark intermediate state.
 Keeping only the top-quark contribution to Eq.$(4.9)$ and taking
 into accout that
\begin{equation}
       \mbox{Im}\{G^q_{jiuk}G^q_{kjiu}\}\sim \frac{g^4}{4\Lambda^4}
                                    \sim \frac{(2\pi)^2}{\Lambda^4}\ ,
\end{equation}
\noindent
we find
\begin{equation}
       d_u=e\frac{m_t}{48\pi^2 \Lambda^2}
            \left[4\ln \frac{\Lambda}{m_t}-2.57\right]\ .
\end{equation}
Since $d_u \gg  d_d$, we find that $d_n = d_u/3$. Assuming that
 $m_t$ = 140 GeV and taking into account the experimental upper
 bound of the neutron electric dipole moment\cite{NEDE} we realize
 that the effective cut-off of the loop integral should satisfy
\begin{equation}
       \Lambda > 800\;  \mbox{TeV}\ .
\end{equation}
\hspace*{\parindent}The above lower bound for the cut-off $\Lambda$
 is of the same order as the one set by the FCNC
 restriction\cite{FCNC}.
 If we use the value of $\Lambda$ set by the FCNC restriction which
 will be described in Eq.$(4.22)$ and calculate $d_n$
 through Eq.$(4.11)$, we find $d_n \sim 5\times 10^{-27}\; e$cm.
 This prediction is surely much smaller than the present experimental
 bound.
 In the standard model with the KM phase the neutron electric dipole
 moment is calculated and is found to be extremely small\cite{NEDM}.
 Our result $(4.11)$ and $(4.12)$ gurantees this property of the
 standard model.
\vglue 0.4cm

{\elevenbf\noindent B. K-meson system}
\vglue 0.3cm
The only known experimental information on the CP violation exists
 in the K-meson decays.
 In this subsection we discuss the $\varepsilon$-parameter which is
 determined by measuring the charge asymmetry in the semileptonic
 decay of the $\mbox{K}^0$ meson and the $2\pi$ decay of the
 $\mbox{K}^0_{L}$ meson.\\
\hspace*{\parindent}$\mbox{K}^0$-meson states
 $|\mbox{K}^{0}_{L}\rangle$ and $|\mbox{K}^{0}_{S}\rangle$ are
 defined as
\begin{equation}
       \left\{
           \begin{array}{c}
           \displaystyle
                  |\mbox{K}^{0}_{L}\rangle
            = \frac{1}{\sqrt{2(1+|\varepsilon|^2)}}
            \left\{(1+\varepsilon)\,|\mbox{K}^0\rangle
            +(1-\varepsilon)\,|\bar{\mbox{K}}^0\rangle
                                      \right\}                  \\
           \displaystyle
                   |\mbox{K}^{0}_{S}\rangle
            = \frac{1}{\sqrt{2(1+|\varepsilon|^2)}}
            \left\{(1+\varepsilon)\,|\mbox{K}^0\rangle
            -(1-\varepsilon)\,|\bar{\mbox{K}}^0\rangle
                                      \right\}
           \end{array}
      \right. \ ,
\end{equation}
\begin{displaymath}
       (\; |\mbox{K}^0\rangle
       = -\, \mbox{\raisebox{-0.1ex}{CP}}\, |\bar{\mbox{K}}^0
                                                \rangle\;  )\ ,
\end{displaymath}
\noindent
With the non-vanishing $\varepsilon$ the K-meson mass eigenstates
 are different from the eigenstates of CP.
 We have
\begin{equation}
       \varepsilon = \frac{\langle\mbox{K}^0|\, H\,
                               |\bar{\mbox{K}}^0\rangle^{\frac{1}{2}}
                              -\langle\bar{\mbox{K}}^0|\, H\,
                               |\mbox{K}^0\rangle^{\frac{1}{2}}}
                              {\langle\mbox{K}^0|\, H\,
                               |\bar{\mbox{K}}^0\rangle^{\frac{1}{2}}
                              +\langle\bar{\mbox{K}}^0|\, H\,
                               |\mbox{K}^0\rangle^{\frac{1}{2}}}
                  \simeq \frac{\langle\mbox{K}^0|\, H'_{viol}\,
                                  |\bar{\mbox{K}}^0\rangle}
                                 {\langle\mbox{K}^0|\, (H_{0}
                                                    +H'_{cons})\,
                                  |\bar{\mbox{K}}^0\rangle}\ .
\end{equation}
Here we require that
\begin{equation}
           |\langle\mbox{K}^0|\, (H_{0}
            +H'_{cons})\, |\bar{\mbox{K}}^0\rangle |
      \gg  |\langle\mbox{K}^0|\, H'_{viol}\,|\bar{\mbox{K}}^0
                                                  \rangle | \ .
\end{equation}
The Hamiltonian $H'_{viol}$ contains the following term,
\begin{equation}
       i\, \mbox{Im}(G)\int \!\!d^{3}\!x\, \bar{s}_{\sL}
           d_{\sR}\bar{s}_{\sR}d_{\sL}+h.c.\ ,
\end{equation}
where $G$ is the corresponding four-fermion coupling constant
 and s and d represent the s and d-quark fields.
Although in our model Im$(G)$ vanishes, we here consider the more
 general cace in which Im$(G)\neq 0$.
Using the PCAC relation (It should be remembered that a specific
 choice of the contraction of color indices is made in Eq.$(3.3)$ ),
 we find
\begin{equation}
       \langle\mbox{K}^0|\bar{s}_{\sL}d_{\sR}
       \bar{s}_{\sR}d_{\sL}|\bar{\mbox{K}}^0\rangle
       =-\frac{B_{K}(\mu )f_{K}^{2}m_{K}^{4}}{2(m_s+m_d)^2}\ ,
\end{equation}
where $B_{K}(\mu)$ is the so-called B-parameter, $f_{K}$ is the
 K-meson decay constant and $m_{K}$, $m_{d}$ and $m_{s}$ are the
 mass of the K-meson, d-quark and s-quark respectively.
 After some calculation, we obtain
\begin{equation}
       \langle\mbox{K}^0|H'_{viol}|\bar{\mbox{K}}^0\rangle
       \simeq -i\, \frac{\mbox{Im}(G)B_{K}(\mu )f_{K}^{2}m_{K}^{3}}
                     {4(m_{d}+m_{s})^2}
              \langle\mbox{K}^0|{\mbox{K}}^0\rangle\ .
\end{equation}
 We see by definition
\begin{equation}
       \langle\mbox{K}^0|\, (H_{0}+H'_{cons})\,
       |\bar{\mbox{K}}^0\rangle
       \simeq \frac{1}{2}\left(\Delta M-\frac{i}{2}\Delta
                                                   \Gamma \right)
              \langle\mbox{K}^0|{\mbox{K}}^0\rangle\ ,
\end{equation}
where $\Delta M$ and $\Delta \Gamma$ are the $\mbox{K}_L-\mbox{K}_S$
 difference of the mass and decay width respectively.
 Accordingly we obtain
\begin{equation}
       \varepsilon \simeq
       \frac{-i\, \mbox{Im}(G)B_{K}(\mu)f_{K}^{2}m_{K}^{3}}
        {2\left(\Delta M-\frac{i}{2}\Delta \Gamma \right)
                                          (m_{d}+m_{s})^2}\ .
\end{equation}
\noindent
By inserting experimental data
 in Eq.$(4.20)$ we find
\begin{equation}
       \mbox{Im}(G)\sim 10^{-9}\; \mbox{TeV}^{-2}\ .
\end{equation}
\noindent
The above result $(4.21)$ is about $10^2$ times smaller than that obtained by
 the FCNC restriction\cite{FCNC},
\begin{equation}
       \mbox{Re}(G)< 10^{-7}\; \mbox{TeV}^{-2}\ .
\end{equation}
\vglue 1.0cm

\stepcounter{subsection}
\stepcounter{subsection}
\stepcounter{subsection}
\stepcounter{subsection}
\subsection{CONCLUSION}
\stepcounter{section}
\vglue 0.1cm
\hspace*{\parindent}Applying Dashen's mechanism to the composite
 Higgs models we succeeded to find simple models of the dynamical
 CP violation.
 Although our models have to be further elaborated to explain
 the actual KM phase,
 they represent an essential ingredient of the dynamical CP
 violation in the standard model and may be thought of as prototype
 models which accommodate the CP violation
 in the standard model.\\
\hspace*{\parindent}In order to see whether our model could be in
 conformity with experimental situations we examined low-energy
 consequences of our model.
 By estimating the $\varepsilon$-parameter in the neutral K-meson
 decays and the neutron electric dipole moment we derived the
 lower bound on the cut-off parameter using the available
 experimental informations.
 The cut-off parameter signals, at the scale determined by the
 low-energy data,
 the existence of the deeper theory for which our model is an
 effective theory.
 The lower bound we obtained is consistent with the one required
 by the constraint on the flavor-changing neutral current.\\
\hspace*{\parindent}Although our model is a simple toy model for
 the dynamical CP violation,
 it may be elaborated to fully account for the CP violation in
 the standard model.
 The investigation in this direction is in progress.
\vglue 1.0cm

\subsection*{ACKNOWLEDGEMENTS}
\vglue 0.1cm
\hspace*{\parindent}The authors would like to thank T.~Onogi
 for discussions and suggestions and K.~Yamawaki for useful
 informations.
\vglue 0.6cm
\newpage

\subsection*{REFERENCES}
\vglue 0.1cm

\vglue 0.5cm
\end{document}